\documentstyle[11pt]{article}
\newcommand{\tr}{\mbox{tr}}
\newcommand{\e}{\mbox{e}}

\renewcommand{\theequation}{\arabic{equation}}
\begin{document}
\bibliographystyle{plain}
\def\m@th{\mathsurround=0pt}
\mathchardef\bracell="0365
\def\upbrall{$\m@th\bracell$}
\def\undertilde#1{\mathop{\vtop{\ialign{##\crcr
    $\hfil\displaystyle{#1}\hfil$\crcr
     \noalign
     {\kern1.5pt\nointerlineskip}
     \upbrall\crcr\noalign{\kern1pt
   }}}}\limits}
\def\theequation{\arabic{section}.\arabic{equation}}
\newcommand{\ar}{\alpha}
\newcommand{\aar}{\bar{a}}
\newcommand{\bb}{\beta}
\newcommand{\gm}{\gamma}
\newcommand{\Gm}{\Gamma}
\newcommand{\en}{\epsilon}
\newcommand{\dd}{\delta}
\newcommand{\sg}{\sigma}
\newcommand{\kp}{\kappa}
\newcommand{\ld}{\lambda}
\newcommand{\oa}{\omega}
\newcommand{\be}{\begin{equation}}
\newcommand{\ee}{\end{equation}}
\newcommand{\bea}{\begin{eqnarray}}
\newcommand{\eea}{\end{eqnarray}}
\newcommand{\bse}{\begin{subequations}}
\newcommand{\ese}{\end{subequations}}
\newcommand{\nn}{\nonumber}
\newcommand{\bR}{\bar{R}}
\newcommand{\bP}{\bar{\Phi}}
\newcommand{\bS}{\bar{S}}
\newcommand{\bU}{\bar{U}}
\newcommand{\bW}{\bar{W}}
\newcommand{\vf}{\varphi}
\newcommand{\sn}{{\rm sn}}
\begin{flushright}
hep-th/9503168
\end{flushright}
\begin{center}
{\Large{\bf New Boundary Conditions for Integrable Lattices $\,{}^1$}}
\vspace{.5cm}

V.B. Kuznetsov\vspace{.3cm} \\
Faculteit voor Wiskunde en Informatica,
Universiteit van Amsterdam$\,{}^{2,3}$\\
Plantage Muidergracht 24, 1018 TV Amsterdam, The Netherlands\vspace{.7cm}\\
M.F. J{\o}rgensen and P.L. Christiansen\vspace{.3cm} \\
Institute of Mathematical Modelling,\\
Technical University of Denmark,\\
DK-2800, Lyngby, Denmark
\end{center}
\footnotetext[1]{Report 95-08, Mathematical preprint series,
University of Amsterdam;  hep-th/9503168.}
\footnotetext[2]{Supported by the Nederlandse Organisatie voor Wetenschappelijk
Onderzoek (NWO).}
\footnotetext[3]{On leave of absence from Department of Mathematical and
Computational Physics, Institute of Physics,
St.~Petersburg University, St.~Petersburg 198904, Russia.}
\vspace{.5cm}
\centerline{\bf Abstract}
\vspace{.2cm}
New boundary conditions for integrable nonlinear lattices of the XXX
type, such as the Heisenberg chain and the Toda lattice
are presented. These integrable extensions are
formulated in terms of a generic XXX Heisenberg magnet interacting
with two additional spins at each end of the chain. The construction uses
the most general rank 1 ansatz for the $2\times 2$ $L$-operator
satisfying the reflection equation algebra with rational $r$-matrix.
The associated quadratic algebra is shown to be the one of dynamical
symmetry for the $A_1$ and $BC_2$ Calogero-Moser problems. Other physical
realizations of our quadratic algebra are also considered.
\vskip 4cm
\pagebreak
%
%
\section{ Introduction}
\setcounter{equation}{0}
\label{intro}
Many of the Liouville integrable lattices, for instance the Toda
lattice and the Heisenberg chain, remain integrable when imposing
some boundary conditions other than the open or the periodic
ones. Usually it corresponds to switching from the $A_{n-1}$
classical root system to the other ones ($B_n$, $C_n$, $D_n$,
$BC_n$, etc.) associated to an integrable lattice \cite{s88,kt92,rs87}.
It was an idea of Sklyanin \cite{s88} (see also \cite{chered})
to describe all such possible
boundary conditions in terms of the representations of a new algebra
\be
R(u-v)T^{(1)}(u)R(u+v)T^{(2)}(v) =
T^{(2)}(v)R(u+v)T^{(1)}(u)R(u-v)\,,
\label{qismii_quant}
\ee
which was called afterwards {\em the reflection equation algebra}
(see for instance \cite{ks92}). We prefer to give it another
name {\em the QISM II algebra} (see \cite{kt92,k90,kk94,k94}), where
QISM stands for the Quantum Inverse Scattering Method and II
symbolizes the difference of it from {\em the QISM I algebra}, which is
nothing but the familiar algebra given by the quadratic relation
\cite{ks92,ft87,b82,s85}
\be
R(u-v)T^{(1)}(u)T^{(2)}(v) = T^{(2)}(v)T^{(1)}(u)R(u-v)\,.
\label{qismi_quant}
\ee
There are actually too many names for these algebras, and no one
of them is becoming standard; that is why we are insisting on our
own choices -- the QISM I and II algebras. These names are short
and quite characteristic too. We should also mention the work \cite{MNO}
where a different version of the reflection equation (\ref{qismii_quant})
was considered.

In terms of the integrable models the representations of the QISM I
algebra~(\ref{qismi_quant}) provide us with integrable lattices (through
the co-multiplication) while the representations of the QISM II
algebra~(\ref{qismii_quant}) describe the possible boundary conditions
for such lattices.
In this paper we use the new representation of the QISM II algebra
found recently in \cite{k94} to obtain some new boundary terms for the
known integrable lattices.
These new models are generically formulated in terms of the XXX
Heisenberg magnet
interacting with two additional spins on each end. We consider as well
the degenerate cases given by the corresponding contraction
procedure.
We would like to stress that our results give for the moment
the most general boundary terms for the Heisenberg magnet (expressed
in terms of two additional spins on each end) and specialization will
give already known boundary conditions \cite{s88,kt92} coming from
the scalar or rank 0 solutions of the QISM II algebra.
Describing the situation in a bit more technical terms we can say
that we are classifying all boundary conditions coming from the most
general rank 1 representations of the QISM II algebra in the XXX case.
All results in this paper are given for the Poisson algebra. The
quantization is straightforward and will be published elsewhere.
%
%
\section{Rank 1 representations of the QISM II algebra}
\setcounter{equation}{0}
\label{reps}
In the classical limit the QISM II algebra~(\ref{qismii_quant}) becomes
the following Poisson algebra \cite{s88}
\bea
\{T^{(1)}(u),T^{(2)}(v)\} &=& [r(u-v), T^{(1)}(u)T^{(2)}(v)]
\label{qismii}\\
&&\mbox{} + T^{(1)}(u)r(u+v)T^{(2)}(v) - T^{(2)}(v)r(u+v)T^{(1)}(u)\,,
\nonumber
\eea
where $T^{(1)}(u)
\equiv T(u)\otimes I$, $T^{(2)}(v) \equiv I \otimes T(v)$.
Here $\otimes$ denotes the usual tensor product and $I$ is the $2\times 2$
identity matrix. The $T(u)$ is the $2\times 2$ monodromy matrix
depending on the complex spectral parameter $u$. The $4\times 4$
matrix $r(u)$ will in this paper be given by
\be
r(u) = \frac{-1}{u}\left(\begin{array}{cccc}
1 & 0 & 0 & 0\\
0 & 0 & 1 & 0\\
0 & 1 & 0 & 0\\
0 & 0 & 0 & 1\end{array}\right).
\label{r}
\ee
This is the rational case, which corresponds to the XXX model
\cite{rs87,ft87}.

Let $T(u)$ be of the following form \cite{kk94,k94}
\be
T(u) = \left(\begin{array}{cc}
A(u) & B(u)\\
C(u) & D(u)\end{array}\right),
\label{tu}
\ee
with
\bea
A(u) &=& \alpha u^2 + A_1 u + A_0 + \frac{\delta}{u}\,,\qquad
D(u)  = -A(-u)\,,
\label{au}\\
B(u) &=& \beta u^2 + B_0\,,\qquad C(u) = \gamma u^2 + C_0\,.
\label{tu_ansatz}
\eea
Here $\alpha$, $\beta$, $\gamma$, and $\delta$ are scalars, while
$A_1$, $A_0$, $B_0$, and $C_0$ are generators of some algebra.
The $T(u)$ (\ref{tu})--(\ref{tu_ansatz})
satisfies the following symmetry property
\be
T(-u) \sim T^{-1}(u)\,.
\ee
Inserting this Ansatz for $T(u)$ into the Poisson algebra~(\ref{qismii})
leads to the following quadratic Poisson algebra $\cal{A}$ for the
generators $A_1$, $A_0$, $B_0$, and $C_0$
\bea
\{A_0,A_1\} &=& \beta C_0 - \gamma B_0\,,\qquad \quad
\{B_0,A_0\} = 2A_1B_0 - 2\beta\delta\,,\label{s}\\
\{B_0,A_1\} &=& 2\alpha B_0 - 2\beta A_0\,,\qquad
\{C_0,A_0\} = -2A_1C_0 + 2\gamma\delta\,,\\
\{C_0,A_1\} &=& 2\gamma A_0 - 2\alpha C_0\,,\qquad
\{C_0,B_0\} = 4A_1A_0 - 4\alpha\delta\,.
\label{subalg}\eea
The determinant of the monodromy matrix $T(u)$
is the generating function for the
center of the QISM II algebra~(\ref{qismii}).
For the Ansatz~(\ref{au})--(\ref{tu_ansatz})
\be
\det T(u) = -(\alpha^2+\beta\gamma)u^4 + {\cal Q}_2 u^2 + {\cal Q}_0 +
  \frac{\delta^2}{u^2}\,,
\label{dettu}
\ee
with
\be
{\cal Q}_2 = A_1^2 - 2\alpha A_0 - \beta C_0 - \gamma B_0
\label{q2}\,,\qquad
{\cal Q}_0 = 2\delta A_1 - A_0^2 - B_0C_0\,.
\label{q0}
\ee
Hence (\ref{q2}) gives two Casimir elements for the
algebra~(\ref{s})--(\ref{subalg}).

The QISM II algebra (\ref{qismii}) with the $r$-matrix~(\ref{r})
admits the following scalar solution \cite{s88,kt92}
\be
K(u) = \left(\begin{array}{cc}
  a + d/u & b\\
  c & -a+d/u\end{array}\right),
\label{k(u)}
\ee
where $a$, $b$, $c$, and $d$ are complex constants.
This is just a special case of~(\ref{tu})--(\ref{tu_ansatz}) with
$\alpha=\beta=\gamma=A_1=0$. Then $A_0$, $B_0$, and $C_0$ all
commute.
We may now combine the two solutions $K(u)$ and $T(u)$ of the QISM II
algebra and define
\be
t(u) = \mbox{tr} \;K^t(-u) T(u)\,.
\label{t(u)}
\ee
Note that $Z: T(u) \mapsto T^t(-u)$ is an automorphism of the
algebra~(\ref{qismii}) with the $r$-matrix (\ref{r}).
It is a property of the QISM II algebra \cite{s88} that
\be
\{t(u),t(v)\} = 0\,.
\ee
Hence the $t(u)$ (\ref{t(u)}) is the generating function for the
integrals of motion of an associated integrable system.
For our special choices of $K(u)$ and $T(u)$ we get
\be
t(u) = (2\alpha a+\beta b+\gamma c)u^2 + H -2d\frac{\delta}{u^2}\,,
\label{trace}
\ee
where
\be
H = 2 a A_0 + b B_0 + c C_0 - 2 d A_1\,.
\label{hamil}
\ee
The Hamiltonian (\ref{hamil}) defines a completely integrable system, since
it is effectively one dimensional. (Subtract two center
elements~(\ref{q2}) from
four generators and divide by two to see that the
algebra~(\ref{s})--(\ref{subalg})
is of one degree of freedom.)
In the Appendix we show how we may eliminate the first term in the
Hamiltonian~(\ref{hamil}) by using the automorphisms of the
quadratic algebra $\cal{A}$. The result is
\be
H = \tilde b B_0 + \tilde c C_0 -2 d A_1\,,
\label{simp_hamil}
\ee
where the new coefficients $\tilde b$ and $\tilde c$ are related
to $a$, $b$, and $c$ through
\be
a^2 + bc = \tilde b\tilde c\label{tilde_b}\,,\qquad
2\alpha a+\beta b+\gamma c = \beta \tilde b + \gamma\tilde c\,.
\label{tilde_c}
\ee
Any Hamiltonian that is a function of $A_1$, $A_0$, $B_0$, and $C_0$
will be completely integrable. However, for this special case --- an arbitrary
linear combination --- we have formulated the problem using the QISM II
algebra. It is
then possible to apply the method of separation of variables, which will be
shown in the Section~\ref{sep}. We now give a physical realization of the
quadratic algebra~(\ref{s})--(\ref{subalg}).
%
%
\section{The o(4) generalized Lagrange top}
\setcounter{equation}{0}
\label{one_top}

The algebra (\ref{s})--(\ref{subalg})
may be embedded into the $\cal{U}({\rm o(4)})$ Lie algebra
with the six generators $J_i$ and $x_i$, $i=1,2,3$, and the Poisson
brackets
\be
\{J_i,J_j\} = \epsilon_{ijk}J_k\,,\qquad
\{J_i,x_j\} = \epsilon_{ijk}x_k\,,\qquad
\{x_i,x_j\} = \epsilon_{ijk}J_k\,.
\label{o4}
\ee
The $\cal{U}({\rm o(4)})$ algebra has the two Casimir elements
\be
{\cal C}_1 = \vec{x}^{\,2} + {\vec{J}}^{\;2}\,,\qquad
{\cal C}_2 = \vec{x}\cdot\vec{J}\,.
\ee
The homomorphism between the $\cal{U}({\rm o(4)})$ algebra with
relations~(\ref{o4})
and the rank 1 QISM II algebra~(\ref{s})--(\ref{subalg})
is given by the following formulas \cite{k94}
\be
A_0 = x_1 J_2 - x_2 J_1\,,\qquad
B_0 = -x_1^2 - x_2^2 - J_3^2\,,\qquad
C_0 = {\vec{J}}^{\;2}\,,\qquad
A_1 = x_3\,,
\label{homomorphism}
\ee
where the scalars $\alpha$, $\beta$, and $\gamma$ appearing in
(\ref{au})--(\ref{tu_ansatz}) are without loss of generality chosen to be
\be
\alpha = 0\,,\qquad \beta  = -1\,,\qquad \gamma = 1\,,\label{alfa}
\ee
with $\delta$ being equal to
\be
\delta = {\cal C}_2 J_3\,.\label{delta}
\ee
Using (\ref{dettu})--(\ref{q0}) we may write the center of the algebra
in terms of the o(4) variables
\be
{\cal Q}_0 = {\cal C}_2^2 + {\cal C}_1 J_3^2\,,\qquad
{\cal Q}_2 = {\cal C}_1 + J_3^2.
\ee
Notice that from this it follows that $J_3$ commutes with all of
$A_1$, $A_0$, $B_0$, and $C_0$.
It is a simple matter to verify
equations~(\ref{s})--(\ref{subalg}) from the
definitions~(\ref{homomorphism}) and~(\ref{o4}).
The integrable Hamiltonian~(\ref{simp_hamil}) becomes
\be
H = -\tilde b {\cal Q}_2 + (\tilde b+\tilde c){\vec{J}}^{\;2}
  + \tilde b x_3^2 - 2 d x_3\,,
\label{ham_phys}
\ee
with $J_3$ as the additional conserved quantity. For the
particular values of $\alpha$, $\beta$, and $\gamma$
in~(\ref{alfa}) it can be shown that $\tilde b$ and
$\tilde c$ are always real when $a$, $b$, and $c$ are.
The Hamiltonian~(\ref{ham_phys}) corresponds to an o(4)
generalized Lagrange top \cite{rs87}.

In the special case $\alpha^2+\beta\gamma=0$, the algebra $\cal{A}$
is degenerate. We then have the following homomorphism to
${\cal U}(\mbox{e}(3))$ \cite{k94}
\be
A_0 = x_1J_2 - x_2J_1\label{e3_start}\,,\qquad
B_0 = -x_1^2 - x_2^2\,,\qquad
C_0 = {\vec{J}}^{\;2}\,,\qquad
A_1 = x_3\,,\label{e3_stop}
\ee
where we have put without loss of generality that
$\alpha=\beta=0$ and $\gamma=1$ with $\delta$ being as in (\ref{delta}).
The algebra generators $J_i$ and $x_i$ satisfy here the Poisson brackets
like in (\ref{o4}) where the last bracket is equal zero (according to
the contraction procedure from o(4) to e(3)). The first Casimir element
${\cal C}_1=\vec{x}^{\,2}$. The Hamiltonian~(\ref{ham_phys}) for this case
corresponds to the Lagrange top in a nonlinear gravity field \cite{rs87}.
%
%
\section{Separation of variables}
\setcounter{equation}{0}
\label{sep}
We now return to the Hamiltonian~(\ref{hamil}) with the Poisson
brackets~(\ref{s})--(\ref{subalg})
and show how this system may be integrated
by the general method of separation of variables applicable to the
QISM II algebra. This method was first applied by Sklyanin in
\cite{s85,s85b} to the representations of the QISM I algebra. See also
\cite{kt92,k90} where the separation of variables was applied for the
first time to the QISM II algebra under the circumstances close to the ones
we have in the present paper.

To do this it is necessary first to apply a similarity transformation
to the matrices $K(u)$ and $T(u)$ in order to make $K(u)$ triangular.
We define the matrix $V$ as
\be
V = \left(\begin{array}{cc}
  a+w & 0\\
  b & 1\end{array}\right),
\ee
where $a$ and $b$ are the parameters in $K(u)$ defined by~(\ref{k(u)})
and $w^2 = a^2 + bc$. Then we introduce $\tilde K(u)$ and $\tilde T(u)$
as
\be
\tilde K(u) = V^t K(u) (V^{-1})^t\,,\qquad
\tilde T(u) = V^{-1} T(u) V
= \left(\begin{array}{cc}\displaystyle
  \tilde A(u)& \tilde B(u)\\
  \displaystyle \tilde C(u)& \displaystyle -\tilde A(-u)
  \end{array}\right).
\ee
The matrices $\tilde K(u)$ and $\tilde T(u)$ are of the same form as
$K(u)$ and $T(u)$, respectively. The $\tilde K(u)$ now looks like
\be
\tilde K(u) = \left(\begin{array}{cc}\displaystyle
  w+\frac{d}{u} & 0\\
  \displaystyle\frac{c}{a+w} & \displaystyle-w+\frac{d}{u}
  \end{array}\right)
= \left(\begin{array}{cc}\displaystyle
  \tilde a+\frac{\tilde d}{u} & \tilde b\\
  \displaystyle\tilde c & \displaystyle-\tilde a+\frac{\tilde d}{u}
  \end{array}\right),
\ee
while we have the following relations between the entries of
$T(u)$ and $\tilde T(u)$
\be
\tilde\alpha = \alpha + \frac{\beta b}{a+w}\,,\qquad
\tilde\beta = \frac{\beta}{a+w}\,,\qquad
\tilde\gamma = -2b \alpha - \frac{\beta b^2}{a+w}  + (a+w)\gamma\,,
\ee
and
\be
A_0 = \tilde A_0 - b \tilde B_0\label{new_a0}\,,\qquad
B_0 = (a+w)\tilde B_0\,,\qquad
C_0 = \frac{2b\tilde A_0 - b^2 \tilde B_0 + \tilde C_0}{a+w}\,,
\label{new_c0}
\ee
where $A_1$ and $\delta$ are not changed.
Since $\det \tilde T(u)=\det T(u)$ we have $\tilde {\cal Q}_2 ={\cal Q}_2$ and
$\tilde {\cal Q}_0 = {\cal Q}_0$.
Furthermore, $\tilde K(u)$ and $\tilde T(u)$ satisfy the QISM II algebra
too, while $t(u)$ defined in (\ref{t(u)}) is unaltered.
Hence, these matrices generate the same integrable system as before.
Because of the triangular form of $\tilde K(u)$ we get
\be
t(u) = \tilde a(\tilde A(u) + \tilde A(-u)) - \frac{\tilde d}{u}
  (\tilde A(u) - \tilde A(-u)) + \tilde c\tilde C(u)\,.
  \label{new_trace}
\ee
Now let $u_1$ be a zero of $\tilde C(u)$, i.e.
\be
\tilde C(u_1) = 0\,,
\label{u_1}
\ee
and define
\be
\lambda_1^+ = -\tilde A(-u_1)\label{l1p}\,,\qquad
\lambda_1^- = \tilde A(u_1)\label{l1m}\,.
\ee
We may easily evaluate
\bea
\lambda_1^+\lambda_1^- &=& -\tilde A(u_1)\tilde A(-u_1)
  = \det \tilde T(u_1) = \Delta(u_1)\,,
\label{prod}
\eea
where $\Delta(u) = \det T(u)$ is given by~(\ref{dettu}),
and from~(\ref{new_trace}) we get
\be
t(u_1) = w(\lambda_1^- - \lambda_1^+) - \frac{d}{u_1}(
  \lambda_1^- + \lambda_1^+)\,.
\label{sum}
\ee
This is nothing but separation equation for the separation variables
$u_1,\lambda_1^\pm$.
Because $\tilde T(u)$ satisfies the QISM II
algebra we have the important Poisson brackets between the new
variables $u_1$ and $\lambda_1^{\pm}$ (see \cite{kt92,k90})
\be
\{u_1,\lambda_1^{\pm}\} = \pm\lambda_1^{\pm}\,,\qquad
\{\lambda_1^+,\lambda_1^-\} = -\Delta'(u_1)\,.
\label{can2}
\ee
{}From~(\ref{prod}) and (\ref{can2}) it follows that we may put
\be
\lambda_1^{\pm} = \sqrt{\Delta(u_1)}\;\e^{\pm p_1}\,,\qquad
\{u_1,p_1\} = 1\,.
\label{l1pm}
\ee
Inserting~(\ref{l1pm}) into (\ref{sum}) and using~(\ref{trace})
gives
\be
H = -2\sqrt{\Delta(u_1)}\left(
w\sinh{p_1}+\frac{d}{u_1}\cosh{p_1}\right)
  - \left(2\alpha a + \beta b + \gamma c\right)
u_1^2 + 2 d\frac{\delta}{u_1^2}\,.\label{new_ham}
\ee
We have thus formulated the problem as an one-dimensional system with the
Hamiltonian~(\ref{new_ham}) in terms of the canonical coordinates
$u_1$ and $p_1$. The equation of motion looks like
\be
\dot u_1 = \{u_1,H\}\,,
\ee
from which we have
\be
(\dot u_1)^2 = t(u_1)^2 - 4\Delta(u_1)\left(\frac{d^2}{u_1^2}-w^2\right).
\ee
After the substitution $v_1=u_1^2$ the problem can be integrated in
terms of elliptic functions.
The transformation from $(A_1,A_0,B_0,C_0)$ to $(u_1,p_1)$ is given by
equations (\ref{hamil}), (\ref{u_1}), and (\ref{new_ham}). To get back
we first write $\tilde A(u)$ and $\tilde C(u)$ in
terms of $u_1$ and $p_1$ through the Lagrange interpolation using the data
\be
\tilde C(\pm u_1) = 0\,,\qquad
\tilde A(\pm u_1) = \pm\lambda_1^{\mp}\,,
\ee
and the leading parameters $\tilde\alpha$, $\tilde\beta$, $\tilde\gamma$,
and $\tilde\delta$.
This gives
\bea
\tilde C(u) &=& \tilde\gamma(u^2-u_1^2)\,,\\
\tilde A(u) &=& \tilde\alpha(u^2-u_1^2) + \frac{\tilde\delta}{u}
  \frac{u_1^2-u^2}{u_1^2} + \frac{u-u_1}{2u_1}\lambda_1^+
  + \frac{u+u_1}{2u_1}\lambda_1^-\,.
\eea
Equating powers of $u$ and using~(\ref{l1pm}) then gives
\be
\tilde C_0 = -\tilde\gamma u_1^2\,,\quad
\tilde A_0 = -\tilde\alpha u_1^2 - \sqrt{\Delta(u_1)}\sinh{p_1}\,,\quad
\tilde A_1 = -\frac{\tilde\delta}{u_1^2} + \frac{\sqrt{\Delta(u_1)}}{u_1}
  \cosh{p_1}\,.
\ee
To get $\tilde B_0$ we insert $\tilde B(u)=\tilde\beta u^2 + \tilde B_0$
into $-\tilde A(u)\tilde A(-u)-\tilde B(u)\tilde C(u)=\Delta(u)$ and equate
powers of $u$. This gives
\be
\tilde B_0 = \frac{1}{\tilde\gamma}\left(
  2 \frac{\tilde\delta^2}{u_1^4} + \frac{\tilde {\cal Q}_0}{u_1^2}
  + \tilde\alpha^2 u_1^2 + \frac{\Delta(u_1)}{u_1^2} \sinh^2{p_1}
+ 2 \sqrt{\Delta(u_1)}\left(\tilde\alpha
  \sinh{p_1}-\frac{\tilde\delta}{u_1^3}\cosh{p_1}\right)\right).
\ee
Alternatively $\tilde B_0$ may be restored from the Poisson
bracket
\be
\{\tilde A_0,\tilde A_1\} = \tilde\beta \tilde C_0 - \tilde\gamma\tilde B_0.
\ee
We may finally get $A_1$, $A_0$, $B_0$, and $C_0$ in terms of $u_1$,
$p_1$, ${\cal Q}_0$, and ${\cal Q}_2$
by using (\ref{new_c0}).
%
%
\section{Two interacting o(4) tops}
\setcounter{equation}{0}
\label{two_tops}
We now generalize the results of the Sections~\ref{reps} and \ref{one_top}
to the case of not just one but two o(4) Lagrange tops
interacting with each other. This is done by replacing in~(\ref{t(u)})
$K(u)$ with $\tilde T(u)$, where $\tilde T(u)$ is of the same form
as $T(u)$, but with a different choice of $\tilde\alpha$,
$\tilde\beta$, $\tilde\gamma$, $\tilde\delta$, $\tilde A_1$,
$\tilde A_0$, $\tilde B_0$, and $\tilde C_0$.
Because the QISM II algebra~(\ref{qismii}) is closed under the action
of any similarity transformation, we may write $t(u)$ as
\be
t(u) = \mbox{tr}\;\tilde T^t(-u)VT(u)V^{-1}\,,
\label{tu_gen}
\ee
where $V$ is any matrix with determinant~1, while $T(u)$ is given
by (\ref{tu})--(\ref{tu_ansatz}) and
(\ref{homomorphism})--(\ref{delta}) and similarly for $\tilde T(u)$.
Putting
\be
V = \left(\begin{array}{cc}
  a & b\\
  c & d\end{array}\right),
\ee
with $ad-bc=1$ leads to the following integrable system
\be
t(u) = (a^2+b^2+c^2+d^2)u^4 - Hu^2 - G - \frac{2\delta\tilde\delta}{u^2}\,,
\ee
where
\bea
H &=& -2(ab+cd)A_0 + (a^2+c^2)B_0 - (b^2+d^2)C_0\nonumber\\
&&\mbox{}-2(ac+bd)\tilde A_0 + (a^2+b^2)\tilde B_0 -
  (c^2+d^2)\tilde C_0 + 2 A_1\tilde A_1\,,\label{H}\\
G &=& -2(1+2bc)A_0\tilde A_0 - a^2 B_0\tilde B_0-d^2C_0\tilde C_0
  +2a(cB_0\tilde A_0+bA_0\tilde B_0)\nonumber\\
&&\mbox{}+b^2C_0\tilde B_0 + c^2 B_0\tilde C_0 - 2 d(bC_0\tilde A_0
  +cA_0\tilde C_0) + 2(\delta\tilde A_1 + \tilde\delta A_1)\,.
\eea
If we regard the $H$ (\ref{H}) as the Hamiltonian then we see that
the interaction between the two systems is through the term $A_1\tilde A_1$.
We may use the automorphisms of the quadratic algebra~$\cal{A}$ to
eliminate the terms in $H$ with $A_0$ and $\tilde A_0$. Using
equations~(\ref{tilde_b}) we get the following Hamiltonian
\be
H = \lambda(C_0-\tilde B_0)+\lambda^{-1}(\tilde C_0-B_0)
  + 2A_1\tilde A_1\,,
\ee
where $\lambda$ is the single nontrivial parameter in the Hamiltonian
satisfying $-(\lambda+\lambda^{-1}) = a^2+b^2+c^2+d^2$. The second
conserved quantity is
\be
G = 2(A_0\tilde A_0 + \delta \tilde A_1 + \tilde\delta A_1)
  - \lambda C_0\tilde B_0 - \lambda^{-1}B_0\tilde C_0\,.\label{fourth}
\ee
Applying the homomorphism~(\ref{homomorphism})
gives the final result
\be
H = {\vec{J}}^{\;2} + {\vec{\tilde x}}^{\,2} + (x_3 \cosh\theta
  + \tilde x_3\sinh\theta)^2\label{new_ham2}\,.
\ee
Here $\lambda=\tanh\theta$, while constant additive terms and
multiplicative factors have been neglected. This is a new integrable system
which we call {\it two interacting o(4) Lagrange tops}\,.
We remark that the number of degrees of freedom is 4 and there are two
simple integrals, namely: $J_3$ and $\tilde J_3$. The fourth integral
$G$ (\ref{fourth}) is of degree 4 in $J_i$'s and $x_i$'s.
No separation of variables is known for this system or in
general for any integrable system
given by (\ref{tu_gen}) in the case when both $T(u)$ and
$\tilde T(u)$ are nonscalar.

The Hamiltonian~(\ref{new_ham2}) contains interesting subcases.
We may easily change the real form speaking about two interacting systems
on the algebras, for instance, o(2,2)$\oplus$o(2,2), o(3,1)$\oplus$o(4)
and so on. The further possibility is the contraction giving the cases like:
e(3)$\oplus$o(4), e(3)$\oplus$o(3,1), e(3)$\oplus$e(3) etc.
Finally, the two o(4) algebras have two o(3) subalgebras by putting
$(x_1,J_2,x_3)=(s_1,s_2,s_3)$ and similarly
$(\tilde x_1,\tilde J_2,\tilde x_3)=(t_2,-t_1,t_3)$. Then
\bea
H &=& \lambda(s_2^2+t_2^2)+\lambda^{-1}(s_1^2+t_1^2)+2s_3t_3\,,\\
G &=& \lambda^{-1}(s_1t_1-\lambda s_2t_2)^2\,.
\eea
This is a special (but non-trivial, one-parameter)
case of the general (two parameter) o(4) Manakov top \cite{kt92,rs87}.
%
%
\section{Further realizations}
\setcounter{equation}{0}
In this Section we consider other homomorphisms of the specialized
quadratic algebra $\cal{A}$ determined by the condition
$\alpha^2+\beta\gamma=0$.
Without loss of generality we may put $\alpha=\beta=0$ and
$\gamma=1$. Applying the techniques of the Section~\ref{two_tops}
leads to the following system
\bea
H &=& C_0+\tilde C_0 + 2\lambda A_1\tilde A_1\,,
\label{spec_ham}\\
G &=& C_0\tilde C_0 + 2\lambda(\delta\tilde A_1 +\tilde\delta A_1 -
  A_0\tilde A_0) + \lambda^2B_0\tilde B_0\,,
\eea
with the single parameter $\lambda$.
We have the following homomorphism of the specialized algebra $\cal{A}$
into ${\cal U}(\mbox{sl}(2)\oplus\mbox{sl}(2))$
\bea
A_0 &=& 4(s_3t_--t_3s_-)\,,\qquad B_0 = -16t_-s_-\,,\\
C_0 &=& (s_++t_+)(s_-+t_-)-(s_3+t_3)^2\,,\qquad A_1 = 2(t_--s_-)\,,\\
\delta &=& 2(s_-+t_-)({\cal C}_s-{\cal C}_t)\,.
\eea
Here ${\cal C}_s$ and ${\cal C}_t$ are the Casimir elements of the
two $\mbox{sl}(2)$ algebras
\be
{\cal C}_s = s_3^2-s_+s_-\,,\qquad
{\cal C}_t = t_3^2-t_+t_-\,,
\ee
and the Poisson brackets for the generators are given by
\bea
\{s_-,s_3\}&=&s_-\,,\qquad \{s_-,s_+\}=2s_3\,,\qquad \{s_3,s_+\}=s_+\,,\\
\{t_-,t_3\}&=&t_-\,,\qquad \;\{t_-,t_+\}=2t_3\,,\;\qquad \{t_3,t_+\}=t_+\,.
\eea
We may realize the pair of $\mbox{sl}(2)$ algebras in terms of
the canonical variables
$(x,y,p_x,p_y)$ via the following homomorphism ($\{x,p_x\}=1$ etc.)
\bea
s_3 &=& -\frac{xp_x}{2}\,,\qquad
s_- = \frac{x^2}{2}\,,\qquad
s_+ = \frac{p_x^2}{2} - \frac{2l}{x^2}\,,\\
t_3 &=& -\frac{yp_y}{2}\,,\qquad
t_- = \frac{y^2}{2}\,,\qquad
t_+ = \frac{p_y^2}{2} - \frac{2m}{y^2}\,,
\eea
where $l$ and $m$ are the values of ${\cal C}_s$ and ${\cal C}_t$,
respectively.
Inserting these relations into~(\ref{spec_ham}) gives the following
Hamiltonian
\bea
H &=& \frac{1}{4}(x p_y-y p_x)^2 + \frac{1}{4}(\tilde x\tilde p_y
  -\tilde y\tilde p_x)^2
  + 2\lambda (x^2-y^2)(\tilde x^2-\tilde y^2)\nonumber\\
&&\qquad \mbox{} - (x^2+y^2)\left(\frac{l}{x^2} + \frac{m}{y^2}\right)
  - (\tilde x^2 + \tilde y^2)\left(\frac{\tilde l}{\tilde x^2}
  + \frac{\tilde m}{\tilde y^2}\right).
\eea
In polar coordinates $x=r\cos\theta,\;y=r\sin\theta,\;\tilde x=
\tilde r\cos\tilde\theta,\;\tilde y=\tilde r\sin\tilde \theta$
this system becomes
\bea
H &=& \frac{1}{4}\;\dot\theta^2 + \frac{1}{4}\;\dot{\tilde\theta}^2
  + 2\lambda r^2 \tilde r^2 \cos 2\theta \cos 2\tilde\theta\nonumber\\
&&\qquad\mbox{} - \frac{l}{\cos^2\theta} - \frac{m}{\sin^2\theta}
  - \frac{\tilde l}{\cos^2\tilde\theta} - \frac{\tilde m}{\sin^2\tilde\theta}
\,,\label{gen_d2}\\
G &=& \frac{1}{16}\left(\dot\theta\dot{\tilde\theta}-2\lambda r^2\tilde r^2
  \cos2(\theta-\tilde\theta) + 2\lambda r^2\tilde r^2
\cos2(\theta+\tilde\theta)\right)^2
  \nonumber\\
&&\qquad\mbox{}+\left(\frac{1}{4}\;\dot\theta^2 - \frac{l}{\cos^2\theta} -
  \frac{m}{\sin^2\theta}\right)\left(\frac{1}{4}\;
  \dot{\tilde\theta}^2 - \frac{\tilde l}{\cos^2\tilde\theta}
  - \frac{\tilde m}{\sin^2\tilde\theta}\right)
-\left(\frac{\dot\theta\dot{\tilde\theta}}{4}\right)^2\nonumber\\
&&\qquad\mbox{}+2\lambda r^2\tilde r^2\left((m-l)\cos2\tilde\theta
  +(\tilde m-\tilde l)\cos2\theta\right)\,.\label{GG}
\eea
Regarding $H$ (\ref{gen_d2}) as the Hamiltonian, we can see that we have
a system of two particles situated in some singular
field and interacting to each other through the product of cosines.
This can be interpreted as two interacting pendulas. This is a new integrable
system. The $r^2$ and ${\tilde r}^{\;2}$ are
constants, being the level values of two simple integrals: $s_-+t_-$ and
${\tilde s}_-+{\tilde t}_-$. The fourth integral $G$ (\ref{GG}) is of degree
4 in terms of $\dot\theta$ and $\dot{\tilde\theta}$. We would like to call
this system {\it the $D_2$ Toda lattice with singular terms}.
Without the singular
terms the system is easily integrated
by introducing $\theta_{\pm} = \theta \pm \tilde\theta$. However,
no integration of the general system is yet known.
In what follows we will give two
more physical examples connected to some realizations of the
specialized algebra ${\cal A}$: namely, the so-called $A_1$ and $BC_2$ rational
Calogero-Moser systems \cite{OP83}.

The $A_1$ rational Calogero-Moser system is given by the Hamiltonian
\be
H = \frac{1}{2}(p_1^2+p_2^2) + \frac{1}{(x_1-x_2)^2}\,,\qquad
\{p_i,x_j\}=\delta_{ij}\,,\label{SUP-H}
\ee
with the additional conserved quantity $P = p_1+p_2$. Introduce three
more variables
\be
R   = x_1+x_2\,,\qquad
H_0 = \frac{1}{2}(p_1x_1 + p_2 x_2)\label{s0}\,,\qquad
H_+ = \frac{1}{2}(x_1^2+x_2^2)\,.\label{splus}
\ee
We now have the following realization of the specialized algebra~$\cal{A}$
\bea
A_0 &=& 4(PH_0-RH)\,,\qquad B_0 = 4(P^2-4H)\,,\qquad C_0 = 4(HH_+-H_0^2)\,,\\
A_1 &=& 2P\,,\qquad \delta = 0\,,
\eea
again with $\alpha=\beta=0$ and $\gamma=1$.
Since ${\cal Q}_2 = 16H$,
$H$ commutes with all of $A_0$, $B_0$, $C_0$, and $A_1$.
This observation corresponds to the fact that the Hamiltonian (\ref{SUP-H})
is superintegrable \cite{W}.

Inserting this realization into the Hamiltonian~(\ref{spec_ham})
for two ``interacting $A_1$ Calogero-Moser systems''
leads to the following integrable system
\bea
H &=& (x_1p_2-x_2p_1)^2 + 2\frac{x_1^2+x_2^2}{(x_1-x_2)^2}
  + (\tilde x_1\tilde p_2-\tilde p_2\tilde x_1)^2  \nonumber\\
&&\qquad\mbox{}+  2 \frac{\tilde x_1^2+\tilde x_2^2}
  {(\tilde x_1-\tilde x_2)^2} +\lambda(p_1+p_2)(\tilde p_1+\tilde p_2)\,.
\eea
By the construction of this system, we know that its motion
on the levels of two Hamiltonians for two $A_1$ Calogero-Moser systems
is equivalent to that of the system~(\ref{gen_d2}) (because both systems
give two different realizations of the {\it same}
specialized algebra ${\cal A}$).

Consider finally the following one-parameter case of the $BC_2$ rational
Calogero-Moser system given by the Hamiltonian
\be
H = \frac{1}{2}(p_1^2+p_2^2)+\frac{1}{(x_1-x_2)^2} +
  \frac{1}{(x_1+x_2)^2} + \frac{\kappa^2}{2}\left(\frac{1}{x_1^2}
  +\frac{1}{x_2^2} \right),
\label{SUP-HH}\ee
where $\kappa$ is a constant.
This system may be embedded into the specialized algebra~$\cal{A}$ in
the following manner
\bea
C_0 &=& \frac{1}{4}(HH_+-H_0^2)\,,\qquad B_0 = 4(N-4H^2)(N-2H^2)\,,\\
A_0 &=& \{N,C_0\}\,,\qquad A_1 = -2(N-3H^2)\,,\qquad
\delta = (\kappa^2-1)H^2\,,
\eea
where $H_0$ and $H_+$ are given by (\ref{splus}) and
$N$ is the so-called second conserved quantity \cite{OP83} ($\{H,N\}=0$)
\bea
N &=& p_1^4 + p_2^4 + 2 \kappa^2\left(
\frac{p_1^2}{x_1^2}+\frac{p_2^2}{x_2^2}\right)
  + 8(p_1^2+p_2^2)\frac{x_1^2+x_2^2}{(x_1^2-x_2^2)^2}\nonumber\\
&&\qquad\mbox{} + 16\frac{x_1x_2p_1p_2}{(x_1^2-x_2^2)^2}
  + \frac{(x_1^4+x_2^4)(\kappa^2(x_1^2-x_2^2)^2 + 4 x_1^2x_2^2)^2}
  {(x_1^2-x_2^2)^4 x_1^4 x_2^4}\,.
\eea
Because the Hamiltonian $H$ is sitting in $\delta$ it commutes with all
the generators and hence the system (\ref{SUP-HH}) is superintegrable too.
In this case the
Hamiltonian~(\ref{spec_ham}) for ``two interacting $BC_2$ Calogero-Moser
systems'' has the following form:
\bea
H &=& (x_1p_2-x_2p_1)^2 + (\tilde x_1\tilde p_2-\tilde x_2
  \tilde p_1)^2 + 2(x_1^2+x_2^2)U(x_1,x_2,\kappa)\nonumber\\
&&\mbox{}+ 2(\tilde x_1^2+\tilde x_2^2)
  U(\tilde x_1,\tilde x_2,\tilde\kappa)
  +\lambda(N-3H^2)(\tilde N - 3\tilde H^2)\,,
\label{beeee}\eea
where
\be
U(x_1,x_2,\kappa) = \frac{1}{(x_1-x_2)^2} + \frac{1}{(x_1+x_2)^2}
  + \frac{\kappa^2}{2}
  \left(\frac{1}{x_1^2} + \frac{1}{x_2^2}\right).
\ee
It is quite interesting to remark that the motion given by the Hamiltonian
(\ref{beeee}) on the fixed levels of two integrals of motion (\ref{SUP-HH})
(one with tilde and one without tilde) is the same as the motion given by the
Hamiltonian of the $D_2$ Toda lattice (\ref{gen_d2}).
The last two results regarding the Calogero-Moser systems were obtained
by one of us in \cite{k94b}.
%
%
\section{Two e(3) tops interacting with the $A_n$ Toda lattice}
\setcounter{equation}{0}
The next generalization we present is that of inserting the Toda
lattice between the two e(3) Lagrange tops (see end of the Section 3).
This is done by using the monodromy matrix for
the Toda lattice
\be
L(u) = L_2(u)\cdots L_{N-1}(u)\,,\label{eee}
\ee
where
\be
L_i(u) = \left(\begin{array}{cc}
  0 & \exp(q_i)\\
  \exp(-q_i) & u - p_i\end{array}\right)\,,\qquad
\{q_i,p_j\} = \delta_{ij}\,.
\label{toda_l}
\ee
The matrix $L(u)$ (\ref{eee}) satisfies the Poisson limit of the QISM I algebra
\cite{ft87,s85b}
\be
\{L^{(1)}(u),L^{(2)}(v)\} = [r(u-v),L^{(1)}(u)L^{(2)}(v)]\,,
\ee
with the same $r$-matrix (\ref{r}) as in~(\ref{qismii}).
Now let $T_1(u)$ and $T_N(u)$ be the representations described in
(\ref{e3_stop}). It is then true that the matrix
\be
T(u) = L(u)T_N(u)L^{-1}(-u)\,,
\ee
satisfies the QISM II algebra (\ref{qismii}) too \cite{s88}. Therefore the
trace
\be
t(u) = {\rm tr}\; T_1^t(-u)L(u)T_N(u)L^{-1}(-u)\,,
\label{tu_toda}
\ee
is a generating function for the integrals of motion for the $A_{n-2}$ Toda
lattice interacting with an e(3) Lagrange top at each end.
The Hamiltonian for this system is as follows:
\be
H = \frac{1}{2}\sum_{i=2}^{N-1}p_i^2 +\vec{J_1}^{\,2} +
  \vec{J_N}^{\,2} - \sum_{i=2}^{N-2}
  \exp(q_{i+1}-q_i) - \e^{q_2}x_{1,3} +
  \e^{q_{N-1}}x_{N,3}\,.
\label{toda_hamil}
\ee
The first three terms describe the kinetic energy of the system, which
is a chain of particles plus two tops at the ends of the chain.
The last three terms are the potentials, the first one being of the form of
Toda-like
interaction between the neighbours in the chain while the two last terms
describe an interaction of the tops with the chain.
It is convenient to choose the following representation of the
specialized quadratic algebra~$\cal{A}$.
Let $\alpha=\beta=0$ and $\gamma=-1$. Then (cf. also \cite{kt92,k90})
\bea
A_1 &=& \cosh q\,,\qquad A_0 = p\sinh{q}\,,\qquad
B_0 = - \sinh^2 q\,,\\
C_0 &=& p^2 + \frac{2c_1}{\sinh^2\frac{q}{2}} + \frac{2c_2}
  {\cosh^2{\frac{q}{2}}}\,, \qquad \{p,q\}=1\,,
\eea
where $c_1$ and $c_2$ are arbitrary constants.
The Hamiltonian~(\ref{toda_hamil}) then becomes
\bea
H &=& \frac{1}{2}\sum_{i=1}^N p_i^2 - \sum_{i=2}^{N-2}
  \exp(q_{i+1}-q_i) - \e^{q_2}\cosh q_1 +
  \e^{q_{N-1}}\cosh q_N\\
 &&\qquad\mbox{} +
  \frac{c_1}{\sinh^2\frac{q_1}{2}} + \frac{c_2}{\cosh^2
    \frac{q_1}{2}} + \frac{c_3}{\sinh^2\frac{q_N}{2}} +
    \frac{c_4}{\cosh^2\frac{q_N}{2}}\,.
\eea
This Hamiltonian was proved to be integrable for the first time
in \cite{i88}. It corresponds to the most general Toda lattice
of the type $D_n$ with four additional singular terms. Now
we have given another proof for its integrability and also shown how
it appears naturally from the combination of the standard $A_{n-2}$
Toda lattice with two tops each interacting to the corresponding
edge particle of the $A_{n-2}$ Toda lattice.
%
%
\section{Two o(4) tops interacting with the Heisenberg magnet}
\setcounter{equation}{0}
Let us consider the XXX Heisenberg magnet. It is given by the following
construction. First, we introduce a chain of
simple $L$-operators \cite{ft87} ($k=1,\ldots,N$)
\be
L_k(u)=1+\frac{i}{u}\pmatrix{s_k^3&s_k^+\cr s_k^-&-s_k^3}
\label{xL}
\ee
each of which satisfies the QISM I $r$-matrix algebra ($T(u)=L_k(u)$)
with the $r$-matrix (\ref{r})
\be
\{T^{(1)}(u),T^{(2)}(v)\} = [r(u-v), T^{(1)}(u)T^{(2)}(v)],
\label{xqismi}
\ee
where the $s$-variables have the following Poisson brackets of the
sl(2) Lie algebra for any $k$:
\be
\{s^3,s^\pm\}=\pm i s^\pm, \qquad \{s^+,s^-\}=2is^3.
\label{xsl2}
\ee
We will also use the real variables $s_k^1$ and $s_k^2$ defined as
$s_k^\pm=s_k^1\mp i s_k^2\,$.
The $L$-operators (\ref{xL}) have the following properties when
conjugate and change the sign of the spectral parameter $u$:
\be
L_k(-u)=\bar L^t_k(u)\,,\qquad \bar L_k(u)=\sigma_2L_k(u)\sigma_2\,,
\qquad u\in{\bf R}\,.
\label{xpr}
\ee
The $L$-operators (\ref{xL}) can be also represented in the form of the
scalar product between two vectors:
\be
L_k(u)=1+\frac{i}{u}(\vec s_k,\vec \sigma),\qquad \vec s_k=(s_k^1,s_k^2,s_k^3),
\label{xform}
\ee
where $\vec\sigma=(\sigma_1,\sigma_2,\sigma_3)$ are the standard
Pauli matrices.

The determinants of the $L$-operators are expressed through the
Casimir elements of the sl(2) algebras:
\be
{\rm det} \;L_k(u)=1+\frac{c_k^2}{u^2},\qquad c_k^2=(s_k^1)^2+(s_k^2)^2+
(s_k^3)^2
\label{xdet}
\ee
and have $\pm i c_k$ as degeneration points.
The monodromy matrix $T(u)=L_N(u)\cdots L_1(u)$ satisfies the algebra
(\ref{xqismi}) too, while its trace
$$
t(u)={\rm tr} \;T(u), \qquad \{t(u),t(v)\}=0
$$
provides us with the complete set of the integrals of motion in involution
for the XXX Heisenberg magnet. There is a well-known rule \cite{ft87}
to write down the local Hamiltonian for the model. Suppose that we have
the homogeneous chain, i.e. all $c_k$ are equal, $c_k=c$. Then one should
calculate the logarithm of the generating function $t(u)$ in the
common point of degeneration of all the $L$-operators, i.e. $u=ic$ in our case.
So, the local Hamiltonian for the XXX Heisenberg chain looks like:
\be
H_{loc}={\rm log} \;|t(ic)|^2=\sum_{k=1}^N {\rm log}\,[2+\frac{2}{c^2}
(\vec s_k,\vec s_{k+1})]\,.
\label{xH}
\ee

Let us now proceed further to introducing the boundary conditions for the
chain with the Hamiltonian (\ref{xH}). First, we pick up a special
representation of the QISM II algebra which was introduced in Sections 2 and 3.
Suppose we have the Lie algebra so(3)$\oplus$so(2,1) given by the following
Poisson brackets for its six generators $s_k$ and $t_k$
\be
\{s_i,s_j\}=\varepsilon_{ijk}s_k,
\label{xalg}
\ee
\be
\{t_1,t_2\}=-t_3\,,\qquad \{t_2,t_3\}=t_1\,,\qquad \{t_3,t_1\}=t_2\,.
\label{xalg1}
\ee
The Casimir elements are
\be
{\cal C}_s=(s_1)^2+(s_2)^2+(s_3)^2=s^2\,,\qquad
{\cal C}_t=-(t_1)^2-(t_2)^2+(t_3)^2=t^2\,.
\label{xCas}
\ee
The following matrix $T(u)$
\bea
T(u)&=&\left(\matrix{
u(s_3-t_3)+2i(s_1t_2-s_2t_1)+\frac{(s_3+t_3)(s^2-t^2)}{u}\cr
i(u^2+s^2+t^2+2s_3t_3+2i(s_1t_1+s_2t_2))}\right.
\label{xT}\\{}\nonumber\\
&&\qquad\qquad\qquad \qquad\left.\matrix{
i(u^2+s^2+t^2+2s_3t_3-2i(s_1t_1+s_2t_2))\cr
u(s_3-t_3)-2i(s_1t_2-s_2t_1)+\frac{(s_3+t_3)(s^2-t^2)}{u}}\right)
\nonumber
\eea
gives a representation of the QISM II algebra (\ref{qismii}).

Let us take the same representation with the tilde-variables $\tilde s_k$
and $\tilde t_k$ which are in direct sum to the spins $\vec s$ and $\vec t$:
\be
\{\tilde s_i,\tilde s_j\}=\varepsilon_{ijk}\tilde s_k,
\label{xalg2}
\ee
\be
\{\tilde t_1,\tilde t_2\}=-\tilde t_3\,,\qquad
\{\tilde t_2,\tilde t_3\}= \tilde t_1\,,\qquad
\{\tilde t_3,\tilde t_1\}= \tilde t_2\,.
\label{xalg3}
\ee
The Casimir elements are
\be
\tilde {\cal C}_s= (\tilde s_1)^2+(\tilde s_2)^2+(\tilde s_3)^2=
\tilde s^2\,,\qquad
\tilde {\cal C}_t=-(\tilde t_1)^2-(\tilde t_2)^2+(\tilde t_3)^2=
\tilde t^2\,.
\label{xCas2}
\ee
The following matrix $\tilde T(u)$
\bea
\tilde T(u)&=&\left(\matrix{
u(\tilde s_3-\tilde t_3)+2i(\tilde s_1\tilde t_2-\tilde s_2\tilde t_1)
+\frac{(\tilde s_3+\tilde t_3)(\tilde s^2-\tilde t^2)}{u}\cr
i(u^2+\tilde s^2+\tilde t^2+2\tilde s_3\tilde t_3
+2i(\tilde s_1\tilde t_1+\tilde s_2\tilde t_2))}\right.
\label{xT2}\\{}\nonumber\\
&&\qquad\qquad\qquad \qquad\left.\matrix{
i(u^2+\tilde s^2+\tilde t^2+2\tilde s_3\tilde t_3
-2i(\tilde s_1\tilde t_1+\tilde s_2\tilde t_2))\cr
u(\tilde s_3-\tilde t_3)-2i(\tilde s_1\tilde t_2-\tilde s_2\tilde t_1)
+\frac{(\tilde s_3+\tilde t_3)(\tilde s^2-\tilde t^2)}{u}}\right)
\nonumber
\eea
gives a representation of the QISM II algebra (\ref{qismii}).

The generating function for the integrals of motion for the XXX
Heisenberg chain with boundary terms looks like
\be
t(u)={\rm tr}\; \tilde T(u)L_N(u)\cdots L_1(u)T(u)L_1(u)\cdots L_N(u).
\label{xtu}
\ee
The local Hamiltonian for the system has the following form:
\bea
H_{loc}&=&{\rm log} \;t(ic)=\sum_{k=1}^{N-1} {\rm log}\,[2+\frac{2}{c^2}
(\vec s_k,\vec s_{k+1})]\label{xxH}\\
&&+{\rm log}\,[c^2(s_3-t_3)-(s_3+t_3)(s^2-t^2)
+s_1^1(-c^2+s^2+t^2+2s_3t_3)\nonumber\\
&&+2s_1^2(s_1t_1+s_2t_2)
+2s_1^3(s_1t_2-s_2t_1)]\nonumber\\
&&+{\rm log}\,[c^2(\tilde s_3-\tilde t_3)-(\tilde s_3+\tilde t_3)
(\tilde s^2-\tilde t^2)
+s_N^1(-c^2+\tilde s^2+\tilde t^2+2\tilde s_3\tilde t_3)\nonumber\\
&&+2s_N^2(\tilde s_1\tilde t_1+\tilde s_2\tilde t_2)
+2s_N^3(\tilde s_1\tilde t_2-\tilde s_2\tilde t_1)]
+{\rm log}\,\left(\frac{-4}{c^2}\right)\,.\nonumber
\eea
This integrable Hamiltonian describes the XXX Heisenberg chain interacting
with two spins $\vec s$ and $\vec t$ at the one end of the chain and
with two spins $\vec{\tilde s}$ and
$\vec{\tilde  t}$ at the other one.
Note that in order
to get the  Hermitean Hamiltonians we had to choose the non-compact
so(2,1) form of the
spins $\vec t$ and $\vec{\tilde t}$. The boundary terms in the
Hamiltonian (\ref{xxH}) generalize the ones in \cite{vega}, which described the
influence of the external magnetic field and, in quantum case,  looked like
(formula (53) in \cite{vega})
\be
H=\sum_{n=1}^{N-1}(\sigma_n^x\sigma_{n+1}^x+\sigma_n^y\sigma_{n+1}^y
+\sigma_n^z\sigma_{n+1}^z)+b_-\sigma_1^z-b_+\sigma_N^z+c_-\sigma_1^-
-c_+\sigma_N^-+d_-\sigma_1^+-d_+\sigma_N^+,
\label{vega}
\ee
where $b_\pm$, $c_\pm$, and $d_\pm$ are constants.
The Hamiltonian like (\ref{vega}) --- with six boundary terms ---
was proved recently to be integrable also for the generalization up to the XYZ
case \cite{japan}.
%
%
\section*{Discussion}
In this paper we constructed several new integrable systems which
appear after imposing boundary conditions on the known integrable
lattices. There were obtained the following results: (a) two interacting
o(4) tops; (b) a new interpretation of the
most general Toda lattice of the $D_N$ type (namely, ``$D_N=A_N\;+$ 2 tops'');
(c) quadratic algebra ${\cal A}$ as a dynamical algebra of hidden symmetries
for
the $A_1$ and $BC_2$ Calogero-Moser problems; (d) the explicit form
of the ``local'' Hamiltonian (\ref{xxH}) for the system which describes
an interaction of the XXX Heisenberg chain with two o(4,{\bf C}) tops.

We believe that our results can be generalized in some possible ways:
first, the quantization which seems quite straightforward,
secondly, the $q$-deformation. There is also an open problem
to integrate new systems or to separate variables for them.

It is also very interesting to understand how many spins
can be added to the Heisenberg chain as some sort of boundary conditons
still having the integrability property. In the present paper we showed
how to add two more spins at each end. Our conjecture is: it is possible to
extend this up to {\it three}
spins at each end of the standard spin chain. This
conjecture came from the study of the hypergeometric orthogonal polynomials
\cite{k94}.
%
%
\section*{Acknowledgments}
One of us (VBK) acknowledges the hospitality of The Technical University
of Denmark through a Guest Professorship.
%
%
\appendix
\section{Automorphisms of the quadratic algebra $\cal{A}$}
\setcounter{equation}{0}
In this Appendix we discuss the automorphisms of the quadratic
algebra ${\cal A}$ given by the relations~(\ref{s})--(\ref{subalg}).
It is a property of the QISM II algebra~(\ref{qismii})
that any similarity transformation of a representation
$T(u)$ is an automorphism of
the QISM II algebra, i.e.
\be
\tilde T(u) = V^{-1} T(u) V\,,
\label{trans}
\ee
for any non-degenerate matrix $V$ satisfies the QISM II algebra too.
Rewriting the Ansatz (\ref{tu})--(\ref{tu_ansatz}) as
\be
T(u) = u^2 \Omega + A_1 u I + X + \frac{\delta}{u}I\,,
\ee
where $I$ is the $2\times 2$ identity matrix, and
\be
\Omega = \left(\begin{array}{cc}
  \alpha & \beta\\
  \gamma & -\alpha\end{array}\right)\,,\qquad
X = \left(\begin{array}{cc}
  A_0 & B_0\\
  C_0 & -A_0\end{array}\right),
\ee
it becomes clear that $A_1$ and $\delta$ do not change under the
transformation~(\ref{trans}).
We wish to find a matrix $V$ that leaves $\alpha$, $\beta$, and $\gamma$
unaltered. This will be so if $V$ commutes with $\Omega$.
We consider therefore the following choice
\be
V(\theta) = \exp\left(\frac{\theta}{\Delta}\Omega\right)\,,
\ee
where $\Delta^2 = \det\Omega = -(\alpha^2+\beta\gamma)$.
The only part of $T(u)$ that is changed is $X$, according to
the following formula
\be
X(\theta) = V(-\theta) X V(\theta)\,.
\label{auto}
\ee
The relation~(\ref{auto}) defines an automorphism of the quadratic
algebra~$\cal{A}$, which leaves $\alpha$, $\beta$, $\gamma$, $\delta$,
and $A_1$ fixed.
An arbitrary linear combination of $A_0$, $B_0$, and $C_0$ may be achieved
by the following formula
\be
F = \tr(AX) = 2a A_0 + b B_0 + c C_0\,,
\label{lin}
\ee
where
\be
A = \left(\begin{array}{cc}
  a & c\\
  b & -a\end{array}\right).
\ee
Substituting $X(\theta)$ for $X$ in~(\ref{lin}) leads to
\be
F = \tr(A(\theta)X)\,,
\ee
where
\be
A(\theta) = V(\theta)AV(-\theta)\,.
\ee
This equation thus gives the result of applying the automorphism~(\ref{auto})
to the linear combination~(\ref{lin}). It is not necessary to know the
explicit form of this equation, but rather the two main properties
\be
\det A(\theta) = \det A\,,\qquad
\tr (A(\theta)\Omega) = \tr(A\Omega)\,.
\ee
The second relation follows from the fact that $V(\theta)$ commutes with
$\Omega$.
Letting $A(\theta)=\tilde A$ we may write the above relations
explicitly as
\be
a^2 + bc = \tilde a^2 + \tilde b \tilde c\,,\qquad
2\alpha a+\beta b+\gamma c = 2\alpha\tilde a + \beta\tilde b +
  \gamma\tilde c\,.
\ee
These relations give two equations to determine the new parameters
$\tilde a$, $\tilde b$, and $\tilde c$ in terms of the old parameters
$a$, $b$, and $c$, in agreement with the additional arbitrary parameter
$\theta$. We may thus choose to put any one of the new parameters to
zero and thus determine the values of the remaining two.

\end{document}